\documentclass[12pt]{article}
\usepackage[cp866]{inputenc}
\usepackage{amssymb}
\usepackage{amsmath}
\usepackage{amsfonts}

\usepackage{xcolor}

\usepackage{graphicx} % пакет для работы со вставкой рисунков
\usepackage{geometry} % пакет для установки размеров страницы

\usepackage{cite}

\begin{document}

\author{M.I. Kalinin}
\title{On the dual nature of a plane angle}
\date{}
\maketitle

\noindent
Research Center for Applied Metrology --- Rostest Moscow, Russia

\noindent
E-mail: mikhailik@rostest.ru

\thispagestyle{empty}%

\begin{abstract}
\noindent %
For decades, metrologists have debated heatedly whether a plane angle is a dimen-sional or dimensionless quantity; whether it is a base quantity in the International System of Units (SI) or a derived quantity. Two main points of view have emerged in the international metrology community. Those who hold the first view believe that a plane angle is a dimensionless derived quantity equal to the ratio of two lengths, and its unit, the radian, is the dimensionless number one (1 rad = 1 m/m = 1). Those who hold the second view believe that a plane angle is a dimensional quantity with its own independent dimension, and its unit, the radian, is not the dimensionless number one, as is currently accepted in the SI. 

This article demonstrates that, depending on the physical situation, a plane angle is described by either a dimensional or a dimensionless quantity. When measuring, expressing, and communicating an angle's size, physicists use the dimensional quantity plane angle. Its dimension and unit are independent of the dimensions of other quantities and their units. This quantity, plane angle, should be classified as a base quantity, and its unit, radian, should be included in the class of base SI units.

In theoretical studies of physical systems with angular quantities, the latter always enter into equations as a dimensionless combination of dimensional plane angles. This dimensionless combination, in turn, is also a physical quantity characterizing the plane angle in question. This new quantity is a dimensionless derived quantity, which physicists also call an angle.

\end{abstract}

{\bf Key words:} plane angle, quantity, unit, dimension, radian.

\section{Introduction. A Brief history of the evolution of angles in the SI}

\label{history}

In 1960, the Eleventh General Conference on Weights and Measures (CGPM) adopted the International System of Units (SI). All SI units were divided into two classes: base units and derived units. Six dimensionally independent units were chosen as base units: the meter, kilogram, second, ampere, degree Kelvin\footnote{In 1967, at the 13th CGPM, the name `kelvin' was adopted for the unit of thermodynamic temperature instead of the name `degree Kelvin'.} , and candela. In 1971, the mole, a unit of amount of substance used in chemistry, was added to the base class and is also considered dimensionally independent of the other base units. All other units must be expressed in terms of the base units using mathematical formulas.

However, on the question of plane and solid angle's units, scientists were unable to develop a unified position regarding which class to assign them to. The radian (rad) and steradian (sr), respectively, were adopted as units of plane and solid angle. They were separated into a third class, Supplementary units. They were considered neither base nor derived units and were not related in any way to other units. Their dimensions in the SI were not defined. In practice, specialists most often considered the radian and steradian dimensional units, independent of each other and all other units.

This undefined status caused misunderstanding and confusion. Since the dimensions of angles were independent of the dimensions of other SI quantities, they were often considered base quantities, and their units were considered base units. The International Committee for Weights and Measures (CIPM) received constant inquiries about the status of these quantities and their units. In 1969, a clarification was provided in a Note to Recommendation 1 of the CIPM \cite{cipm-58_1969}: "The term 'supplementary units' used in Resolution 12 of the Eleventh General Conference on Weights and Measures (and in this Recommendation) applies to SI units for which the General Conference has not decided whether they are base units or derived units." It is clear that such an explanation did not clarify anything for anyone.

In 1979, the Technical Committee on Quantities and Units of the International Organization for Standardization (ISO TC 12) addressed the CIPM through the Consultative Committee on Units (CCU) with a request \cite[P. U14, Document CCU/80-3]{ccu-7_1980}: ``\ldots to take the necessary actions to resolve the question of whether the radian and steradian are derived or base units.'' This address presented arguments in favor of recognizing plane and solid angles as dimensionless derived quantities, and their units, the radian and steradian, as the dimensionless number one: 1 rad = 1 m/m = 1, 1 sr = 1 m$^2$/m$^2$ = 1.

In 1980, at its 7th meeting, the CCU supported the position of ISO TC 12 regarding the SI supplementary units \cite[P. U12, Recommendation U1]{ccu-7_1980}. In the same year, the CIPM supported the CCU recommendation and decided \cite{cipm-69_1980} to interpret the class of Supplementary units in the SI as a class of dimensionless derived units. And since 1981, in the SI Brochure, the radian and steradian, while formally remaining in the class of supplementary units, are considered dimensionless derived units. It was only in 1995 that the 20th CGPM finally approved this decision, eliminating the class of Supplementary units.

However, debates over the dimensions and status of angles and their units continue to this day. Moreover, the discussion of angular quantities is expanding and intensifying, encompassing the frequencies of periodic processes, the angular frequen-cy, and the phase angle of oscillatory processes. These issues are discussed at meetings of ISO TC 12, CCU, CIPM, and other organizations. And so far it has not been possible to come to a consensus.

There are two main points of view regarding the status of the plane angle in the international metrology community.

{\it The first point of view.} A plane angle is a dimensionless derived quantity, and its unit, the radian, is a dimensionless number one, 1 rad = 1 m/m = 1. This follows from the definition of the quantity plane angle in \cite[P. U15, Document CCU/80-6]{ccu-7_1980}:

``In ISO 31/1, the angle is defined by a statement which is equivalent to the equation (in obvious notations) $\theta = s/r$. Defined in this way the angle is a derived quantity of dimension: ${\dim}\, \theta = {\mathrm L}\cdot{\mathrm L}^{-1}$, i.e. the quantity is dimensionless.''

In this definition, $r$ is the radius of a circle with its center at the vertex of the geometric figure `angle', $s$ is the length of the arc of the said circle enclosed between the two sides of the angle, $\theta$ is the defined physical quantity `angle', $\mathrm L$ is the dimension of the physical quantity length.

This point of view is the official position of the CGPM.

{\it The second point of view.} A plane angle is a quantity with its own dimension, independent of the dimensions of other quantities. Its unit, the radian, is not a dimensionless number and should be included in the class of SI base units. The circle of scientists supporting this point of view has recently expanded significantly.

The next section presents arguments supporting the independent, non-zero di-mension of the plane angle. Section 3 demonstrates that the plane angle enters into physics equations as a dimensionless combination, which physicists also call an angle. This dimensionless combination (dimensionless angle) is a derived quantity in the SI. The formula adopted by the ISO for defining the plane angle is also explained. The article's conclusion proposes including the dimensional plane angle in the class of base quantities, and its unit, the radian, in the class of SI base units.

\section{Dimensional plane angle}

\label{dimension}

The quantity plane angle should characterize a physical object, which is also called plane angle\footnote{In literature, a plane angle (both a physical object and a quantity) is often simply called an angle for brevity}. We'll take the definition of this object from a mathematical encyclopedia \cite{Vinogradov-5}: ``An angle is a geometric figure consisting of two distinct rays emanating from a single point. The rays are called the sides of the angle, and their common origin is the vertex of the angle.''

How is the physical quantity a plane angle defined? What property of a given physical object does it characterize? The mathematical encyclopedia \cite{Vinogradov-1} states that ``Each specific kind of quantity is associated with a specific method of comparing physical bodies or other objects. For example, in geometry, segments are compared by superposition, and this comparison leads to the concept of length.'' Thus, the dimensional quantity length arises. 

\begin{figure}  %[h]
\center
\includegraphics[width=0.5\linewidth]{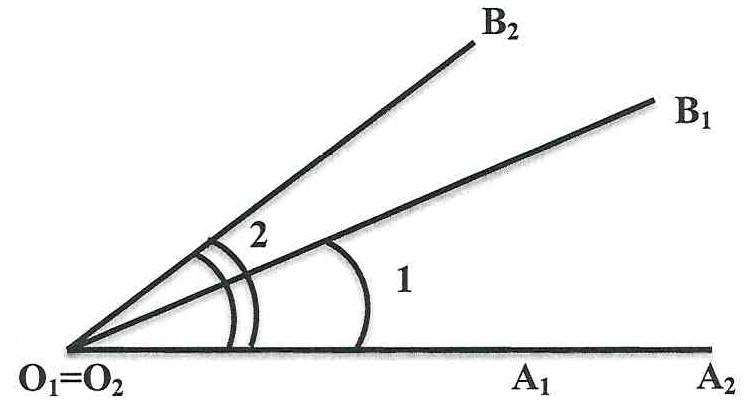}
\caption{Comparison of plane angles}
\label{comparison}
\end{figure}

Similarly, using the superposition method \cite{ Vinogradov-5} to compare angles, we can arrive at the concept of the dimensional quantity plane angle (see Fig. \ref{comparison}).

Two angles are called equal (or congruent) if they can be combined so that their vertices and corresponding sides coincide. Comparison of unequal plane angles is performed as follows \cite{ Vinogradov-5}. The angles are placed in the same plane and the vertices of angles $\mathrm O_1$ and $\mathrm O_2$ and one pair of their sides  $\mathrm O_1\mathrm A_1$ and   $\mathrm O_2\mathrm A_2$ are combined so that both angles are located on one side. If the second side $\mathrm O_1\mathrm B_1$ of the angle $\mathrm A_1\mathrm O_1\mathrm B_1$ is located inside another angle $\mathrm A_2\mathrm O_2\mathrm B_2$, then the first angle $\mathrm A_1\mathrm O_1\mathrm B_1$ is smaller than the second $\mathrm A_2\mathrm O_2\mathrm B_2$. And you can even say how much smaller -- at the angle $\mathrm B_1\mathrm O_1\mathrm B_2$.

It's clear from the figure itself that the plane angle, denoted by the symbol $\varphi$, is the degree of deviation of one side of the angle from the other. This deviation does not depend on the length of the sides of the angle or on any other physical objects. This means that the plane angle, $\varphi$, does not depend on any other quantities. Consequently, the plane angle cannot be a derived quantity, and should be considered the eighth base quantity in the SI with its own independent dimension, which can be denoted by the symbol A (see, for example, \cite{Eder, Leonard}).

If the directions of the two sides of an angle coincide, then we assume that the value of such a plane angle is zero $\varphi = 0$. When one side of the angle rotates, the degree of deviation of its sides from each other, that is, the quantity $\varphi$, will increase from zero to a value corresponding to one complete revolution, when the directions of both sides of the angle again coincide. We denote this value by $\Phi$.

Further rotation of the side of a geometric angle does not lead to a further increase in $\varphi$, but to a cyclical repetition of the process described above. That is, the quantity plane angle, which characterizes the geometric object plane angle, can take values within a finite interval $0 \le \varphi < \Phi$. None of the current seven base SI units possesses this property. Formally they can all take values from zero to infinity.

It is clear that an angle of a specific size should be chosen as the unit of meas-urement for a plane angle. The natural basis for defining a unit of measurement for a plane angle is the quantity $\Phi$. Essentially, $\Phi$ is a natural constant associated with the geometry of flat two-dimensional space. This constant has always been used as the basis for defining a particular unit of plane angle. Historically, the plane angle was the first quantity whose unit was defined by fixing the exact value of the defining constant, long before the creation of the International System of Units. Even in ancient Babylon, the unit of measurement for plane angle, the degree, was defined by the equality $\Phi = 360^{\circ}$. Today, all known units of plane angle are also defined by this constant: $\Phi=360^{\circ} = 2\pi\,\mathrm{rad} = 400\,\mathrm{grad} = 1\, \mathrm{rev} = 32\, \mathrm{rhumbs}$.

Thus, we conclude that the plane angle is a dimensional quantity independent of other quantities. Therefore, it should be considered a base quantity in the SI. The radian is adopted as the unit of plane angle in the SI. Since the radian (like other units of plane angle) is independent of other units, it should be included in the class of SI base units. This proposal has been made by many authors before (see, for example, \cite{Eder, Leonard, Wittmann, Emerson, Kalinin-1, Kalinin-2, Quincey, Kalinin-3, Kalinin-4}).

The quantity plane angle is used to describe not only the geometric figure of a plane angle, but also the process of rotation. In this case, the plane angle quantity, which characterizes rotation, is not limited to the interval $(0, \Phi)$ specified above. Rotation can continue indefinitely and in different directions, leading to the concept of positive and negative plane angles in an infinite interval $-\infty<\varphi<\infty$. All symbols here denote dimensional angle values, not dimensionless numbers.

The above conclusion about the non-zero dimension of the plane angle quantity is consistent with the existing criterion for the dimensionality of quantities in physics \cite{Sedov}: ``Quantities whose numerical value depends on the choice of units of measurement are called dimensional quantities. Quantities whose numerical value does not depend on the choice of units of measurement are called dimensionless quantities.'' For example, mass, energy, and velocity are dimensional quantities. The fine structure constant, however, is a dimensionless quantity for which even the form of the physical formula depends on the choice of units, but nevertheless retains its numerical value in all systems of units. According to this definition, a plane angle is a dimensional quantity, just like mass and length.

\section{Dimensionless angle}

\label{nondimension}

Why was plane angle declared a dimensionless derived quantity in the SI, and its unit, the radian, a dimensionless number one? The fact is that ISO TC 12, based on Standard ISO 31, proposed the mathematical formula $\theta=s/r$  as the definition of a quantity plane angle. There is no justification (either physical or mathematical) for this definition of plane angle in either ISO standards or any other metrology guidelines.

This formula is well known to mathematicians. It was derived to determine the relationship between the values of the length of an arc of a circle, its radius and the value of a plane angle. Its derivation made no prior assumptions about the dimension of the plane angle $\varphi$. Only the well-known property of circular geometry -- the proportionality of the ratios of central angles and the corresponding circular arcs -- was used. From this, the following relationship is easily derived (see, for example, \cite{Leonard, Kalinin-1, Kalinin-2, Kalinin-3, Kalinin-4}): $$\varphi/\Phi = s/2\pi r.$$

Introducing the notation  
\begin{equation}
\label{teta=fi/Fi}
\theta=\varphi/(\Phi/2\pi),
\end{equation}
we will get the same formula 
\begin{equation}
\label{teta=s/r}
\theta=s/r. 
\end{equation}
As we see, here $\theta$ is not the dimensional plane angle defined above in Section2, but a dimensionless combination of two dimensional plane angles $\varphi$ and $\Phi/2\pi$. The transformations given above did not impose any restrictions on the dimension of $\varphi$.

 Formula~(\ref{teta=s/r}) is {\bf not a definition of the quantity} plane angle, but rather a {\bf determination of the numerical value} of the plane angle, expressed in radians, through the values of the arc length and radius of a circle. That is, the derivation of formula~(\ref{teta=s/r}) is valid provided that the radian is already defined as the dimensional quantity plane angle of a given size  $\Phi/2\pi$.
 
 The dimensionless quantity (\ref{teta=fi/Fi}) in physics and mathematics is also called an angle, like the quantity $\varphi$. In metrology, the quantity (\ref{teta=fi/Fi}) is called the numerical value of the plane angle, expressed in radians. Clearly, the quantity $\theta$, defined in this way, is a dimensionless derived quantity in the SI. However, it is not a derived of the quantity length, as stated in the SI Brochure, but of the plane angle.

Let us point out one more consideration in favor of the definitions of a dimensional plane angle and a dimensionless angle defined by formula~(\ref{teta=fi/Fi}). As Emerson pointed out  \cite{Emerson}, an object plane angle is a local object, defined at a single point, at which both the vertex of the angle and the two directions of the sides of the angle are specified. It is clear that the definition of the dimensional quantity plane angle as the degree of deviation of one side of the angle from the other is also local. As for the dimensionless angle $\theta$, definition (\ref{teta=fi/Fi}), as the ratio of two dimensional plane angles  $\varphi$ and $\Phi/2\pi$, defined at a single point, is local. Formula~(\ref{teta=s/r}) defines a non-local quantity -- the ratio of the length of a circle's arc to its radius. This is despite the fact that the circle itself has no relation to the local geometric object plane angle.

As a result, it turns out that one physical object, a plane angle, is described in different situations by two different quantities with different dimensions. In what situations is the dimensional plane angle $\varphi$ used, and in what situations is the dimensionless quantity (\ref{teta=fi/Fi}) used? Many authors (see, for example, \cite{Eder, Leonard, Emerson, Quincey} have pointed out that the measurement of an angle, its expression, and the transmission of its size are possible only with the dimensional quantity $\varphi$. On the other hand, in theoretical physics, the quantity plane angle always enters into equations in the form of dimensionless combination (\ref{teta=fi/Fi}). The first example is the appearance of quantity (\ref{teta=fi/Fi}) in the derivation of formula ~(\ref{teta=s/r}), presented above.

As a second example, let's consider the parameters of rotation. Let a point body rotate in a plane at a distance $r$ from the center. We characterize the rotation by the change in the dimensional angle of rotation over time,  $\varphi(t)$. In the SI, the first derivative of this angle with respect to time, $d\varphi/dt$,  is called the angular velocity, and the second, $d^2\varphi/dt^2$, is called the angular acceleration. Accordingly, in the SI, the unit of angular velocity is rad/s, and the unit of angular acceleration is rad/s$^2$.

Now let's relate the linear velocity of a rotating point to its angular velocity, and the centripetal acceleration to the angular acceleration. By definition, the linear velocity of a rotating point $v$ is equal in magnitude to the time derivative of the arc length of a circle $v=ds/dt$ and is directed along the tangent to the circle along which the point moves. The centripetal acceleration $a$ is directed along the radius toward the center of rotation and is equal in magnitude to the second time derivative  $a=d^2s/dt^2$. The units of linear velocity and acceleration will be m/s and m/s$^2$, respectively. Now, recalling relations (\ref{teta=fi/Fi}) and (\ref{teta=s/r}), we can write
\begin{equation}
\label{v-a}
v=rd\theta/dt,      \qquad            a=rd^2\theta/dt^2,
\end{equation}
where $\theta$ is defined by formula~(\ref{teta=fi/Fi}).

The first and second derivatives of the quantity   $\theta$ with respect to time are not referred to as angular velocity and angular acceleration in metrological terminology. However, following a proposal by Leonhard Euler \cite{Euler}, physicists call the quantity $d\theta/dt$ angular velocity, and $d^2\theta/dt^2$ angular acceleration. Physicists and mathematicians refer to the quantity $\theta$ itself, as an angle, as mentioned above.

In metrology terminology, the quantity $d\theta/dt$  should probably be called the ``rate of change of the numerical value of a plane angle, expressed in radians,'' and $d^2\theta/dt^2$ should be called the ``the acceleration of change of the numerical value of a plane angle, expressed in radians.'' However, such cumbersome constructions are unsuitable for use as working terms, especially given the enormous number of physical systems in which angular quantities must be described. Therefore, physicists use the same terms for both the dimensional angle $\varphi$ and the dimensionless combination $\theta$, as well as their time derivatives. 

It is precisely this duality in the nature of the physical object plane angle that is the main cause of the serious confusion that has arisen and continues to grow in the metrology of angular quantities. Although physicists and mathematicians generally manage to resolve this confusion in their own work, for most people it presents a serious problem. This problem is related not only to physics and mathematics, but also to linguistic peculiarities, habits, and traditions \cite{Bulygin}.

\section{Conclusion}

The analysis conducted in this article points to the dual nature of the quantity plane angle. Depending on the tasks in which the angle is used, it can be characterized as either a dimensional or a dimensionless quantity. When it is necessary to measure an angle, communicate its size to others, or compare the sizes of different angles, the dimensional quantity plane angle is used. This quantity determines the degree of deviation of one side of a geometric angle from the other. This deviation is independent of the length of the angle's sides or other physical quantities. Consequently, the dimensional quantity plane angle $\varphi$ should be considered a base quantity in the SI with its own independent dimension. Accordingly, the unit of the dimensional angle $\varphi$ is not the dimensionless number one, but a plane angle equal to $\phi/2\pi$, which is called a radian. Since it is independent of any other SI units, it must be included in the class of SI base units.

In theoretical descriptions of physical systems and processes involving angles, the dimensional plane angle $\varphi$ always enters into the equations of physics as a dimensionless combination. This means that all solutions to such equations will depend on the angle $\varphi$ only through the specified dimensionless combination $\theta$.
 
 Therefore, in the SI, there are two distinct quantities with different dimensions that characterize the physical object plane angle. The first quantity is the dimensional plane angle $\varphi$, independent of all other quantities. The second is a dimensionless derived quantity, which physicists also call an angle. The dimensional plane angle should be considered a base SI quantity, while the dimensionless angle is a derived quantity. It should be emphasized that the dimensionless derived quantity (\ref{teta=fi/Fi}) proposed here differs from the dimensionless quantity (\ref{teta=s/r}), which is currently the official SI definition of the quantity plane angle.
 
 Thus, in physics, two angular quantities of different dimensions are used equally: the dimensional plane angle $\varphi$ and the dimensionless angle $\theta=\phi/(\Phi/2\pi)$. Both of these quantities are important in the study of various physical systems.
 
 The unit of a dimensional plane angle is not the dimensionless number one, but an angle equal to one radian. This unit should be included in the class of SI base units.

\end{document}